\documentclass[10pt,twocolumn,aps,prl,superscriptaddress,floatfix, groupedaddress]{revtex4-1}

\pdfoutput=1
\usepackage{amsmath,amsthm,amssymb}
\usepackage[utf8]{inputenc}
\usepackage{mathtext}
\usepackage{times}
\usepackage{epsfig}
\usepackage{amsfonts}
\usepackage{amsmath}
\usepackage{amssymb}
\usepackage{color}
\usepackage{multirow}
\usepackage[normalem]{ulem}
\usepackage{latexsym}
\usepackage{amsfonts}
\usepackage{mathrsfs}
\usepackage{natbib}
\usepackage{verbatim}
\usepackage{graphicx}
\usepackage{xcolor}

\begin{document}

\title{Energy-Time Entangled Two-Photon Molecular Absorption}
\author{D.~Tabakaev, M.~Montagnese, G.~Haack, L.~Bonacina, J.-P.~Wolf, H.~Zbinden, R.~T.~Thew}
\affiliation{D\'epartement de Physique Appliqu\'ee, Universit\'e de Gen\`eve, 1211 Gen\`eve, Switzerland}

\begin{abstract} 
Nonlinear spectroscopy and microscopy techniques are ubiquitous in a wide range of applications across physics and biology. However, these usually rely on high-powered pulsed laser systems. A promising alternative is to exploit entangled two-photon absorption (ETPA), which can lead to tens of orders of magnitude lower incident flux rates than in conventional two-photon absorption (TPA) schemes. However, the role of different entangled degrees of freedom in ETPA was unclear following recent experimental studies, when compared to earlier theoretical works. Here, we first demonstrate a linear dependence of the ETPA rate with the photon-pair flux, a clear signature of ETPA, and estimate the first values for the concentration-dependent ETPA cross-section for Rhodamine 6G. We then investigate the signature of energy-time entanglement and polarization dependence in the ETPA fluorescence rate and demonstrate a strong dependence of the signal on the inter-photon delay that reflects the coherence time of the entangled two-photon wave-packet. 

 
\end{abstract}

\maketitle

Building on Marie G\"{o}ppert-Mayer's discovery of two photon absorption (TPA), non-linear spectroscopic techniques have become invaluable tools for both fundamental and applied research, providing the opportunity to study atomic and molecular transition levels that would be unattainable with linear spectroscopy~\cite{boyd2003nonlinear}. However, these techniques typically use relatively high peak power pico- or femto-second pulsed lasers to compensate for the low probability of two photons arriving simultaneously at the same atom or molecule \cite{bebb1966multiphoton, mollow1968two}. 

\begin{figure*}
	\centering
	\includegraphics[width=0.7\paperwidth]{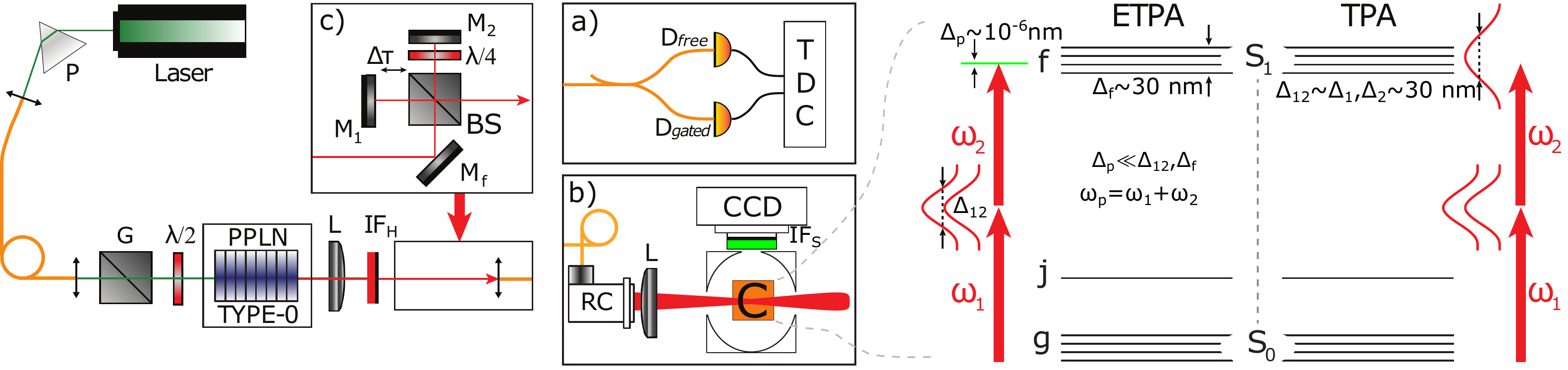}
\caption{Experimental Schematic: A 532\,nm laser beam [Coherent Verdi V5] was sent through a prism (P) and coupled to a single-mode fiber. A Glan prism (G) and waveplate ($\lambda/2$) were used to ensure linearity of the pump polarization before a 2\,cm Type-0 quasi-phase-matched PPLN crystal, producing SPDC-pairs. To remove any residual 532 nm pump photons, a set of high-pass interference filters ($\text{IF}_{\text{H}}$) [Thorlabs FELH0750] was used. The photon coincidence counting setup (box a) consists of a 50:50 fiber beamsplitter, free-running single-photon detector ($D_{\text{\textit{free}}}$) [ID Quantique ID220], gated single-photon detector ($D_\text{\textit{gated}}$) [ID Quantique ID201] and a time-to-digital converter (TDC) [ID Quantique ID801]. In box b, the pairs were coupled to a fiber, passed through a reflective collimator (RC) and focused into an integrating sphere [Thorlabs IS236A-4] containing a custom-made cuvette filled with the Rh6G solution. The counts were collected by a CCD camera (Atik 383L+) attached to one of the integrating sphere's ports. A short-pass filter ($\text{IF}_{\text{S}}$) [Thorlabs FESH0650] was installed before the camera to further reduce spurious detection events. Pairs sent to the interferometer (box c) were separated on the beamsplitter (BS), to introduce either a delay ($\Delta\tau$), or a polarization rotation ($\lambda/4$ waveplate) between the photons. The figure on the right represents the energy level diagram of the absorbing specimen and the various bandwidths for ETPA and TPA regimes. } \label{fig:Schematic} 
\end{figure*}

A promising solution to address this photo-sensitive limitation is to exploit the concept of entangled two-photon absorption (ETPA)~\cite{fei1997entanglement}. Consider a simple model for two photon absorption~\cite{fei1997entanglement} where one photon is excited from the ground state (g) to a virtual level and the other photon further excites this to the final state (f), as illustrated in the energy level diagram on the right of Fig.~\ref{fig:Schematic}. In the case of classical TPA, the absorption rate of two photons is a product of two independent single-photon absorption rates, resulting in a quadratic dependence on the photon flux density $\phi$, $R_{c} =\delta_r \phi^{2} \, [1/\text{s}]$~\cite{fei1997entanglement, dayan2007theory, schlawin2017entangled}. Here, $\delta_r$ is the classical TPA cross-section in units of $\text{cm}^4$\,s. However, if the photons are produced in the form of entangled pairs, they act more like a single object, resulting in a linear absorption rate, $R_{e}=\sigma_{e}\phi \, [1/\text{s}]$, where $\phi$ is now the photon-pair flux density and $\sigma_{e} [\text{cm}^2]$ is the ETPA cross-section~\cite{fei1997entanglement}. We can then write the overall two-photon absorption rate as $R_2 = \delta_r \phi^{2}  + \sigma_{e}\phi$. 

The linear dependence of the absorption rate on the photon-pair flux is a signature that the process is due to ETPA ~\cite{fei1997entanglement,dorfman2016nonlinear,villabona2017entangled,lee2006entangled,lee2007quantum, harpham2009thiophene,upton2013optically,varnavski2017entangled,schlawin2017entangled}.
The ETPA process dominates at low flux density, before the classical, quadratic, TPA takes over at $\tilde{\phi}= \sigma_{e}/\delta_r$. To give an idea of the advantage provided by ETPA in terms of the required flux densities, the typical values for classical TPA are around $\delta_r \sim10^{-47}$\,cm$^4s$~\cite{sperber1986s}, while for ETPA, values as high as $\sigma_{e}\sim10^{-17}$\,cm$^2$ have been obtained~\cite{lee2006entangled,harpham2009thiophene, villabona2017entangled}. 

Motivated by this, there have been numerous theoretical studies to develop spectroscopic techniques based on ETPA to investigate a wide range of molecular systems~\cite{dorfman2016nonlinear}. If we consider the energy-level diagram on the right of Fig.~\ref{fig:Schematic}, in the case of TPA, the transition from the ground state is due to absorption of two photons with bandwidths $\Delta_1$ and $\Delta_2$, which are comparable in value to the bandwidth of the final state $\Delta_f$ (for Rh6G). In contrast, due to the energy-time entanglement, the ETPA transition behaves like it is induced by a spontaneous parametric down conversion (SPDC) pump photon with bandwidth $\Delta_p \ll \Delta_f$, similarly to single photon absorption (SPA).

Recently, several experiments emphasized the role of polarization entanglement~\cite{villabona2017entangled, lee2006entangled, harpham2009thiophene, varnavski2017entangled, lee2007quantum, upton2013optically, guzman2010spatial, guzman2010organic}, while the original theoretical works~\cite{fei1997entanglement, dayan2007theory} and other experiments \cite{dayan2004two,dayan2005nonlinear, pe2005temporal} focused on energy-time entanglement~\cite{Franson89}. In this work we look to clarify this situation by analyzing the ETPA-induced fluorescence intensity of Rh6G as a function of different degrees of freedom of the pairs. This choice for the molecule is motivated by strong absorption properties of Rh6G in the desired wavelength range 510-540\,nm and its high quantum yield \cite{kubin1982fluorescence}. The measurements we performed allow us to demonstrate that the correlations due to energy-time entanglement provide the fundamental advantage for ETPA .

{\it Experimental setup --} The experimental setup is based on a periodically-poled Lithium niobate (PPLN) Type-0 SPDC source for generating entangled photon pairs and three different elements as shown in Fig.~\ref{fig:Schematic}, Boxes a-c). This type of source allows us to avoid any ambiguity about the nature of ETPA by minimizing the entangled degrees of freedom. As we will show, it allowed us to observe the dependence of ETPA-induced fluorescence rate as a function of the photon-pair flux, as well as the relative signal-idler temporal-delay and polarizations. This photon-pair source also allows us to improve the photon-pair flux, and hence to have a greater measurement dynamic compared to previous studies \cite{lee2006entangled,lee2007quantum,varnavski2017entangled}.

Calibration of the pair-source is done by sending the pairs to a standard coincidence detection scheme comprised of two single-photon detectors and a time-to-digital converter; box a). We measured an average coincidence detection rate of $1100$\,s$^{-1}$ at 5.1\,mW of pump power. Taking into account losses, the beamsplitter, and the detection efficiencies of around $3\%$ at 1064\,nm, this corresponds to around $2.6\times10^6 $\,s$^{-1}$ fiber-coupled pairs. This calibration was made in a low flux regime to avoid saturating the detectors. 

Once the calibration was done, the SPDC pairs were sent to the fluorescence detection setup (Fig. 1, box b), where they were focused (60$\mu$m waist) into the cuvette containing the Rh6G molecules in an ethanol solution. The cuvette was placed inside a two-inch integrating sphere, with a (17.6\,mm x 13.52\,mm sensor) CCD-camera attached to one of its half-inch ports, to detect the fluorescence induced by SPDC-pairs and maximize the collection of fluorescence photons. Another port of the sphere was left open, opposite the input, to allow the unabsorbed pairs to pass through - this minimized the chances of spurious detection events due to residual 1064 nm photons, which is further reduced by a short-pass filter installed before the camera.

The role of energy-time entanglement and polarization dependence in the ETPA-induced fluorescence was investigated by inserting a Michelson interferometer before the fluorescence detection setup as shown in box c). In one arm of the interferometer a variable path-length can be controlled to introduce a delay between the photons, and hence, the arrival time difference of the photons sent to the cuvette. Additionally, a $\lambda /4$ waveplate is introduced in one path of the interferometer to rotate the polarization of one photon with respect to the other to test any polarization dependence of the input state. 

\begin{figure} [t]
	\centering
	\includegraphics[width=0.95\linewidth]{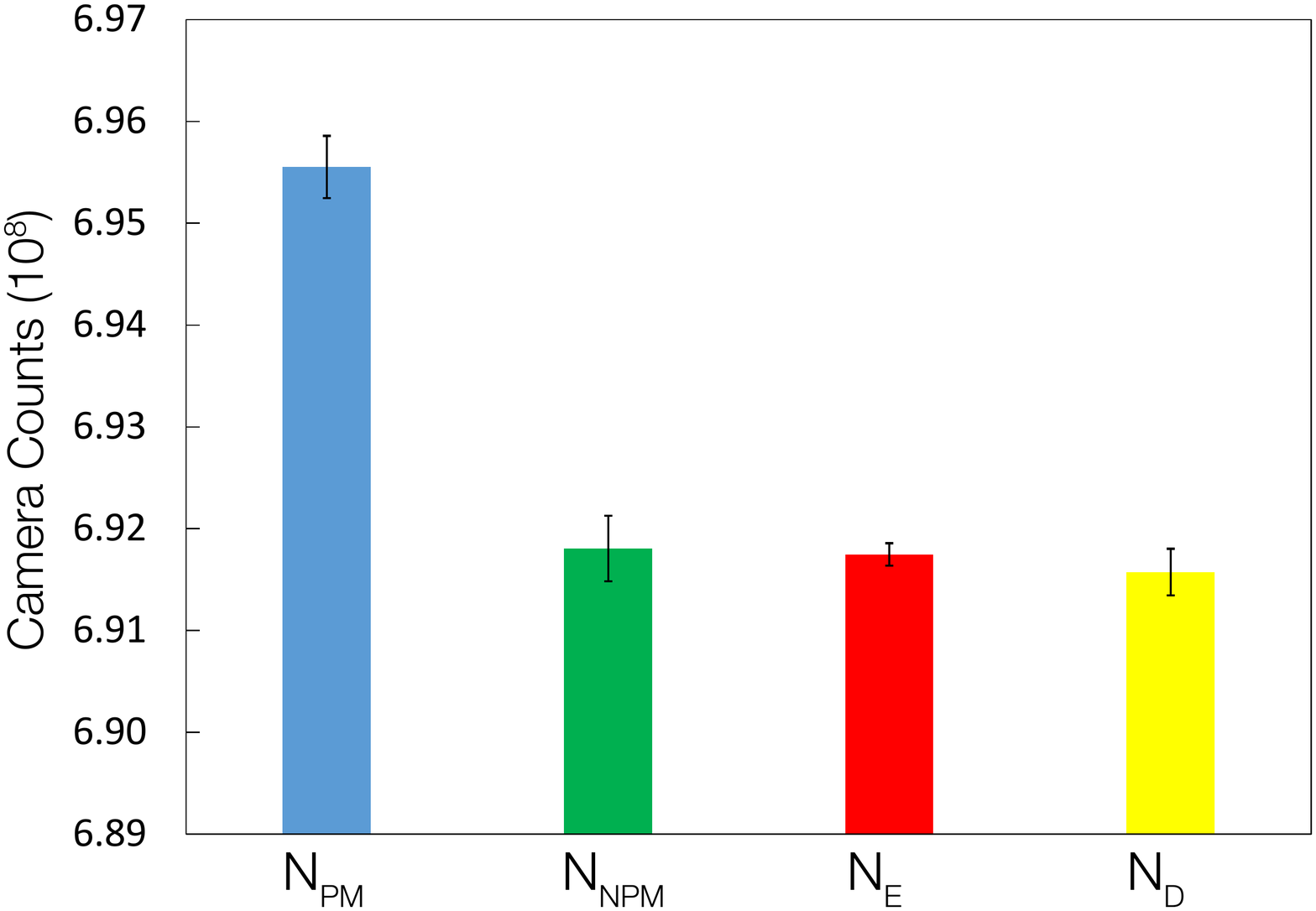}
	\caption{Contributions to the detected photon count rate signal. $N_{PM}$: Raw ETPA-induced fluorescence for a 110 mmol/l Rh6G ethanol solution with 1W of pump power. $N_{NPM}$: Non-phase matching condition - PPLN crystal temperature decreased below that for degenerate SPDC. $N_{E}$: Cuvette with pure ethanol -  no Rh6G in cuvette. $N_{D}$: The laser was turned off and there were no photons incident on the sample. Each result is an average of 10 measurements of 300\,s.}
	\label{fig:RawDataComparison}
\end{figure}

{\it ETPA linearity and cross-section --}
The measurement of the ETPA-induced fluorescence dependence on the photon pair flux density represents the first step to demonstrate the linear signature of ETPA and its dependence on entangled properties of the pair. The combination of an integrating sphere and camera was used to improve the collection efficiency, however, the large camera sensor area also introduces a high background noise rate. Measurements at such low levels of flux and a relatively low signal-to-noise ratio require careful calibration of the system. To achieve this, a series of measurements was made to quantify all contributions to the detected signal (see Fig.~\ref{fig:RawDataComparison} for typical values). To obtain the true signal we extracted the background camera counts from measurements with a sample of pure ethanol ($N_{E}$) from those with Rh6G ($N_{PM}$). These measurements were repeated for each data run and concentration. Importantly, we also ensured that there were no events due to leakage of the  532\,nm pump ($N_{NPM}$), which, due to its high single-photon-absorption cross-section would also produce a linear response to pump power. This was achieved by tuning the temperature away from the phase-matching condition, such that no pairs were produced. $N_{D}$ corresponds to the camera's dark counts when the laser was turned off. These tests confirmed that the detected signal is only due to the fluorescence of Rh6G induced by 1064\,nm photon-pairs.

\begin{figure}[t]
	\centering
	\includegraphics[width=0.9\linewidth]{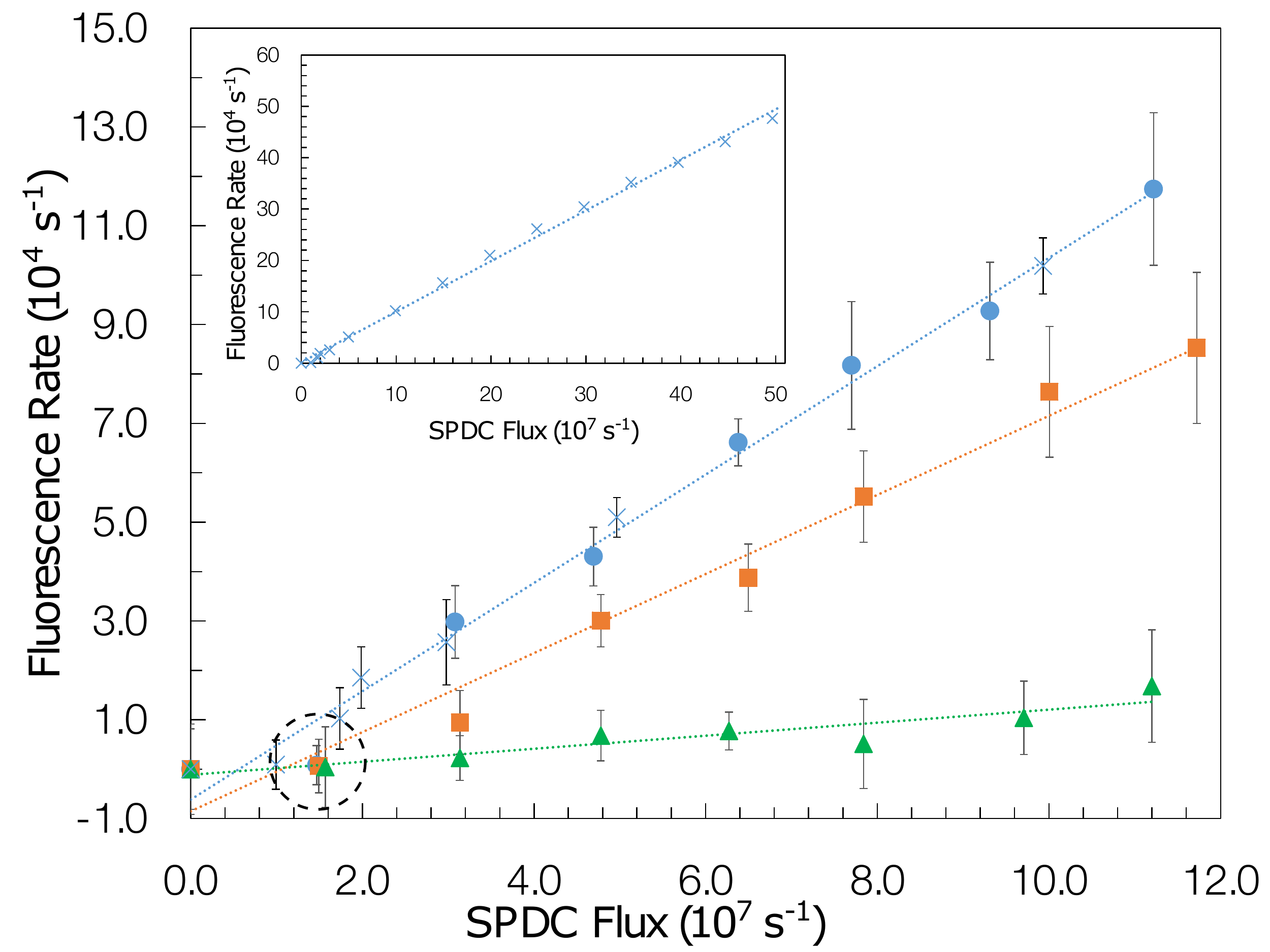}
	\caption{ETPA-induced fluorescence rate as a function of the SPDC photon-pair flux for different Rh6G concentrations in ethanol: 110\,mmol/l (blue circles); 4.5\,mmol/l (orange squares), and 38\,$\mu$mol/l (green triangles). A supplementary measurement (blue crosses) uses a free-space set-up at 110\,mmol/l allowing for higher flux rates (inset). Each point is an average of 10 measurements of 300\,s with the background subtracted.}
	\label{fig:Concentration}
\end{figure}

Figure~\ref{fig:Concentration} shows the measurements of the fluorescence $R_\text{fl}$ rate for three different Rh6G concentrations, $C = \{ 110 \, \text{mmol/l},  4.5 \,\text{mmol/l}, 38 \, \mu\text{mol/l} \}$ as a function of the number of SPDC-pairs sent to the sample. These are the first ETPA absorption measurements with Rh6G and clearly demonstrate a linear dependence of the fluorescence rate on the photon-pair flux density. The pump power was varied over 0.25-1.5\,W, which corresponds to $10^8 - 10^9$ pairs per second sent to the Rh6G solution through the fiber. However, taking into account the dispersion in the fiber (2\,m, connecting source and sample) and the width of the down-converted spectrum, only 10\,\% of the overall flux, corresponding to 1063-1065\,nm range, arrived within the (140\,fs) coherence time of the SPDC two-photon wave-packet. Thus, the effective photon-pair flux incident to the Rh6G solution varies from $2.0\times10^{7}$ to $1.2\times10^{8}\,s^{-1}$. 
To determine the ETPA cross-section we need to connect the fluorescence and absorption rates. The measured fluorescence rate $R_\text{fl}$ can be described by:
\begin{equation}
\label{eq:Rfl}
R_\text{fl} = \frac{N_{PM} - N_{E}}{t_\text{exp}} \frac{G}{\eta_\text{coll} \, \eta_\text{cam}} 
\end{equation}
where $t_\text{exp}$ is the integration time and the fluorescence rate is weighted by the gain factor of the camera $G$, the collection efficiency $\eta_{\rm col}$ and the camera quantum efficiency $\eta_{\rm cam}$. 

Assuming the photon-pair flux is undepleted, the ETPA and fluorescence rates are related through the quantum yield $Y$ that is concentration- and solvent-dependent~\cite{villabona2017entangled}:
\begin{equation}
\label{eq:abs_rate}
R_\text{fl}  = YR_\text{abs} = Y \, C \, V\, N_A\,\sigma_e\,\phi\,.
\end{equation}
Here, $C \, [\text{mol/l}]$ is the concentration, $V=5.6\times 10^{-9}\,l$ is the active volume, $N_A$ is Avogadro's number and $\sigma_e \phi = R_e$ is the "single-molecule" ETPA absorption rate as discussed in the introduction.

In Eq.~\eqref{eq:Rfl}, the factors $G$, $\eta_{col}$ and $\eta_{cam}$ are difficult to quantify individually. We therefore determined them from a relative measurement with the 532\,nm pump laser for SPA and estimate them all together. To do this, we sent the 532\,nm pump laser through the system and replaced the high-pass filters $\text{IF}_{\text{H}}$ with a short-pass filter (Thorlabs FESH0650). The pump beam was attenuated to SPDC-intensity levels, such that $10^8$ photons per second were focused in the cuvette. We then use Eq.~\eqref{eq:Rfl}, Eq.~\eqref{eq:abs_rate} and our knowledge of the SPA cross-section to determine $ G / (\eta_\text{coll} \, \eta_\text{cam})= 4.5\pm0.9 \, \text{(counts)}^{-1}$. 
\begin{table}[!h]
	\begin{tabular}{cc}
		$C$ [mmol/l] \hspace{1cm} & $\sigma_e$ [cm$^{2}$] \\ \hline
		4.5 \hspace{1cm}  & $(9.9\pm 4.9)\times10^{-22}$ \hspace{1cm} \\
		0.038 \hspace{1cm}  & $(1.9\pm 0.9)\times10^{-21}$ \hspace{1cm} \\
	\end{tabular}
	\caption{Results for the ETPA cross-section for two different Rh6G concentrations.}\label{tab:CrossSection}
\end{table} 

Table~\ref{tab:CrossSection} shows the values obtained for $\sigma_e$ for low concentrations, where the quantum yield $Y= 0.95$ is known for Rh6G in an ethanol solution~\cite{kubin1982fluorescence}. This would suggest values of $\sigma_e \sim 10^{-21}$\,cm$^2$, which can be compared to SPA cross-sections of around $10^{-17}$\,cm$^2$ and TPA of around $10^{-47}$\,cm$^4$\,s~\cite{sperber1986s}. A value of the yield for concentrations of 110 \,mmol/l in ethanol is not available, to the best of our knowledge, so we can only report the product, $Y\sigma_e=(6.4\pm 0.5)\times10^{-23}$\,cm$^2$.

This clearly demonstrates a significant quantum advantage with respect to the necessary photon flux rates compared to the TPA cross-section. While these, and the following interferometric, measurements were more practical in fibre, we performed one series in free-space. These results match well with the fiber-based measurements but also provide a higher flux rate, underlining the linear response (crossed points in Fig.\ref{fig:Concentration} and inset).

\begin{figure}[t]
	\centering
	\includegraphics[width=0.9\linewidth]{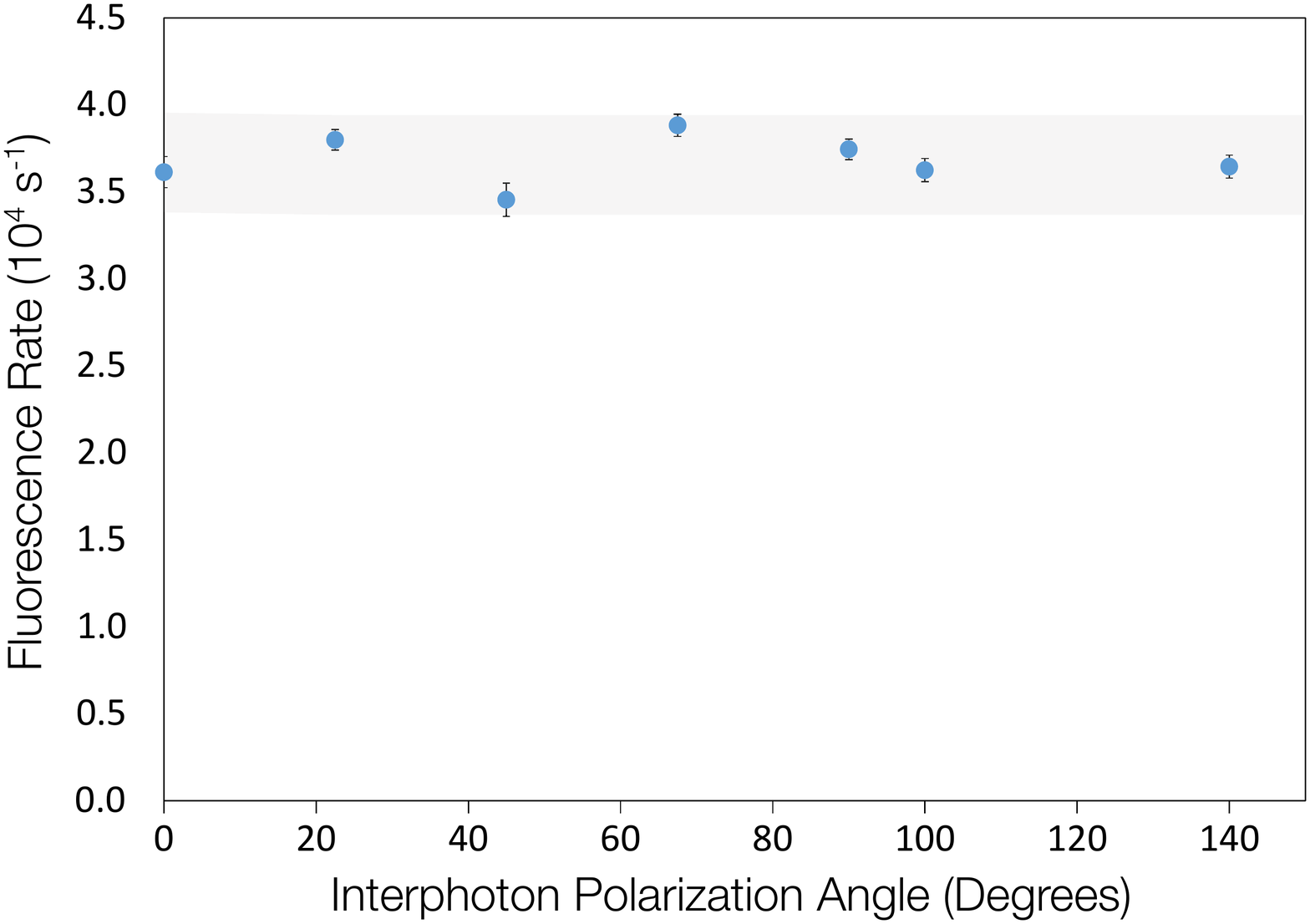}
	\caption{Polarization dependence of ETPA-induced fluorescence rate for a 110 \,mmol/l Rh6G ethanol solution and an SPDC flux of $4.2\times10^7$ pairs/s as a function of the relative angle between the polarization of photons exiting from each arm. Each point is an average of 15 measurements of 300\,s with the background subtracted.} \label{fig:Polarization}
\end{figure}

{\it ETPA polarization and temporal dependence --} To investigate the dependence of ETPA-induced fluorescence on the polarization of the pairs, we rotated the $\lambda$/4 wave-plate in one of the interferometer arms (Fig.~\ref{fig:Schematic} box c) so we change the two-photon state from $|HH\rangle$ to a mixture of states $\{HH, HV, VV, VH\}$. In other words, we scanned between both photons having the same polarization to photons having orthogonal polarizations. Figure~\ref{fig:Polarization} shows the resulting fluorescence rate as a function of the polarization angle where the values vary by less than 1\%.

\begin{figure}[!t]
	\centering
	\includegraphics[width=0.9\linewidth]{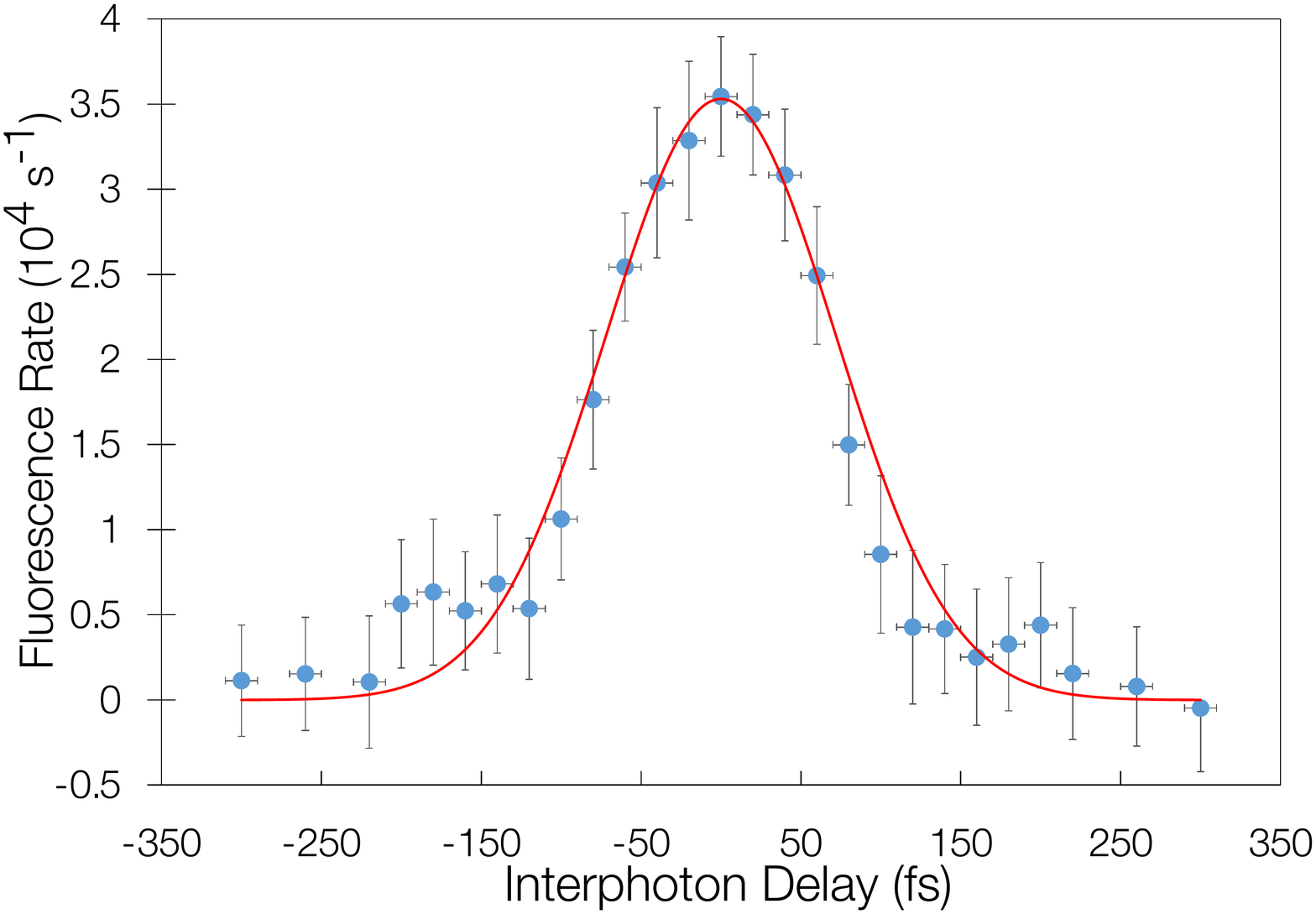}
	\caption{Temporal dependence of ETPA-induced fluorescence rate for a 110mmol/l Rh6G ethanol solution and an SPDC flux of $4.2\times10^7$ pairs/s, as a function of the inter-photon delay $\Delta\tau$. The  interferometer arm was scanned in 20\,fs steps and each point is an average of 15 measurements of 300\,s with the background subtracted. The solid line corresponds to a Gaussian fit with a  FWHM $\sim$\,140\,fs.}
	\label{fig:Delay}
\end{figure}

A simple way to test the role of energy-time correlations of the photons arriving to the unit of absorbing media on the ETPA-rate is to vary the time delay between the photons on the scale of their coherence time (see Fig.~\ref{fig:Schematic}, box c). Figure~\ref{fig:Delay} shows the fluorescence rate as a function of the time delay between photons from an entangled pair for a fixed Rh6G concentration and photon-pair flux. The solid line is a Gaussian fit with a full width at half maximum of approximately 140\,fs, which corresponds to the coherence time of the photon pairs. The small non-zero experimental background is due to photons that take the same path in the interferometer and hence do not have any time delay dependence, and the ``shoulders'' are from pairs that have their (fiber) dispersion cancelled by the time delay. In principle this background should be half the maximum, however, here it drops almost to the noise level. This drop is due to a lower than expected fluorescence signal at low input flux rates (especially for high concentrations); see the circled region in Fig.~\ref{fig:Concentration}. This drop is consistently reproducible but further study is required to understand its origins. 

{\it Conclusion --} We have performed a detailed study of ETPA in a molecular solution of Rh6G in ethanol and extracted the first values for the ETPA cross-section, which shows a concentration-dependent response similar to~\cite{villabona2017entangled}. We demonstrated the main signature of ETPA, i.e. a linear dependence of the absorption rate with the photon-pair flux. We also demonstrated a strong dependence of the signal on the inter-photon delay that reflects the coherence time of the entangled two-photon wave-packet. This can be seen as a fs-sensitive coincidence scheme, analogous to a Hong-Ou-Mandel experiment~\cite{hong1987measurement},  revealing the form of the wavepacket. However, there was no apparent dependence on the polarization of the photons probing the sample, indicating that it is the energy-time entanglement that plays a fundamental role in ETPA.

Besides its fundamental interest, this work also demonstrates the maturity of quantum technologies, such as SPDC sources, for ETPA studies. This will be of particular interest for sensing in general~\cite{nasiri2019entangled}, for photo-sensitive samples, in-vivo studies and microscopy, or alternatively extending recent studies of coherent control for isomerization~\cite{prokhorenko06, prokhorenko2011coherent, palczewska14} to the entangled photon pair regime. There is also a growing interest in studying the vision process, where recent studies have reported classical TPA in humans~\cite{palczewska14} or even as a novel technique for studying the human vision process and whether humans can see single or entangled photons~\cite{Vivoli16,Tinsley16,Holmes18}.

{\it Acknowledgements --} We acknowledge support from the Swiss National Science Foundation through the Sinergia grant CRSII5-170981 and G.~H. acknowledges support through the PRIMA starting grant PR00P$2\_1$79748.

\bibliography{ETPA}

\end{document}